\documentclass[letterpaper,12pt]{article}
\usepackage{amssymb,amsmath,wasysym}
\usepackage{graphicx}
\usepackage[lofdepth,lotdepth]{subfig}
\usepackage[hmargin=1cm,vmargin=1cm]{geometry}
\usepackage{setspace}
\usepackage{float}
\usepackage{subfloat}
\usepackage{titling}
\usepackage{blindtext}
\usepackage{wrapfig}
\usepackage{fancyhdr}
\usepackage{gantt}
\usepackage{pdflscape}
\bibstyle{pra}
\usepackage{floatflt}
\usepackage{pdfpages} 

\renewcommand{\thispagestyle}[1]{}

\begin{document}
\pagenumbering{gobble}


# Astro2020 Science White Paper

# Unlocking the Capabilities of Future High-Resolution X-ray Spectroscopy Missions Through Laboratory Astrophysics

**Thematic Areas:**  ☒ Galaxy Evolution  ☒ Cosmology and Fundamental Physics

**Principal Author:**
Name: Gabriele Betancourt-Martinez
Institution: Institut de Recherche en Astrophysique et Planétologie (IRAP)/CNRS
Email: gabriele.betancourt@irap.omp.eu
Phone: +33 5 61 55 75 32

**Co-authors:** (names and institutions; * marked for contributing author) Hiroki Akamatsu (SRON), Didier Barret (IRAP/CNRS), Manuel Bautista* (WMU), Sven Bernitt (Friedrich-Schiller-Universität Jena), Stefano Bianchi* (Università degli Studi Roma Tre), Dennis Bodewits (Auburn Univ.), Nancy Brickhouse (SAO), Gregory V. Brown* (LLNL), Elisa Costantini (SRON), Marcello Coreno (ISM-CNR, Istituto Struttura della Materia), José R. Crespo López-Urrutia* (MPIK), Renata Cumbee (NASA/GSFC), Megan Eckart (LLNL), Gary Ferland (Univ. of Kentucky), Fabrizio Fiore (INAF), Michael Fogle (Auburn Univ.), Adam Foster* (SAO), Javier Garcia (Caltech), Tom Gorczyca (WMU), Victoria Grinberg (Eberhard Karls Universität Tübingen), Nicolas Grosso (Aix-Marseille Univ./CNRS/CNES/LAM), Liyi Gu* (RIKEN), Ming Feng Gu* (Space Science Laboratory, UC Berkeley), Matteo Guainazzi (ESA), Natalie Hell* (LLNL), Jan-Willem den Herder (SRON), Jelle Kaastra (SRON), Timothy Kallman* (NASA/GSFC), Julia Lee (Harvard Univ.), Maurice Leutenegger* (NASA/GSFC), Joan Marler (Clemson Univ.), Dan McCammon (Univ. of Wisconsin, Madison), Shinya Nakashima (RIKEN), Fabrizio Nicastro (INAF), Frits Paerels* (Columbia Univ.), François Pajot (IRAP/CNRS), Etienne Pointecouteau* (IRAP/CNRS), Delphine Porquet (Aix-Marseille Univ./CNRS/CNES/LAM), F. Scott Porter (NASA/GSFC), Daniel Wolf Savin (Columbia Univ.), Makoto Sawada (NASA/GSFC), Chintan Shah* (MPIK), Aurora Simionescu (SRON), Monica de Simone (IOM-CNR, Istituto Officina dei Materiali), Chad Sosolik (Clemson Univ.), Phillip Stancil (Univ. of Georgia), René Steinbrügge* (Deutsches Elektronen-Synchrotron DESY), Hiroya Yamaguchi (JAXA)

**Abstract:** Thanks to high-resolution and non-dispersive spectrometers onboard future X-ray missions such as XRISM and Athena, we are finally poised to answer important questions about the formation and evolution of galaxies and large-scale structure. However, we currently lack an adequate understanding of many atomic processes behind the spectral features we will soon observe. Large error bars on parameters as critical as transition energies and atomic cross sections can lead to unacceptable uncertainties in the calculations of e.g., elemental abundance, velocity, and temperature. Unless we address these issues, we risk limiting the full scientific potential of these missions. Laboratory astrophysics, which comprises theoretical and experimental studies of the underlying physics behind observable astrophysical processes, is therefore central to the success of these missions.



\pagestyle{fancy}
\lhead{Astro2020 White Paper}
\rhead{Laboratory Astrophysics for X-ray Spectroscopy}
\renewcommand{\headrulewidth}{0.4pt}

\normalsize
\clearpage
\pagenumbering{arabic}

\vspace{2ex}

Spectroscopic observations in the X-ray band hold keys to exciting new discoveries about the origin and nature of the universe. In the past decade, thanks to high-resolution spectrometers onboard missions like \textit{Chandra} \cite{2005PASP..117.1144C}, \textit{XMM-Newton} \cite{2000SPIE.4012..102D}, and \textit{Hitomi} \cite{2016SPIE.9905E..0VK}, we have advanced our understanding of plasma cooling at the centers of galaxy clusters \cite{2003ApJ...590..207P}, measured the first high-resolution spectra of active galactic nuclei (AGN) winds \cite{2000A&A...354L..83K, 2000ApJ...535L..17K}, and learned that the core region of the Perseus cluster of galaxies, often expected to be turbulent, is surprisingly calm \cite{2016Natur.535..117H}. These studies, along with hundreds of others made possible by current and past X-ray spectroscopy missions, push the boundaries of our understanding of galaxy formation, galaxy evolution, and cosmology.


\begin{floatingfigure}[r]{3.35in}
\includegraphics[totalheight=2.2in]{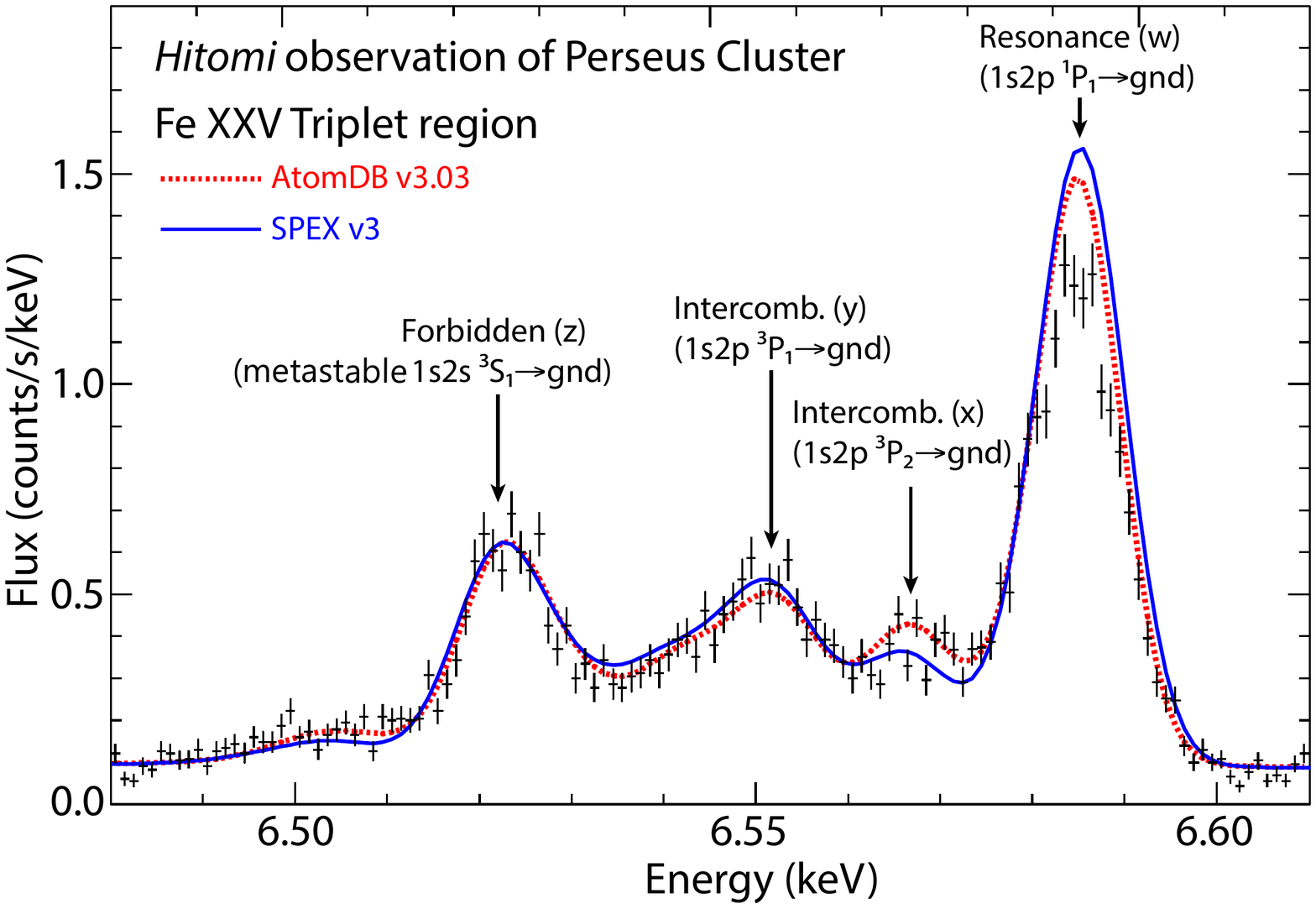}
\caption{\it The He-like line ``triplet" is often used as an electron density and temperature diagnostic in the X-ray regime, but relevant plasma models can disagree at the 10\% level both with each other and the data. The observed lower flux of the ``w" line may be due to resonant scattering or, at least in part, to inaccurate models.}
\end{floatingfigure}

While this work has been groundbreaking, our ability to extend our knowledge further is, in many cases, limited by the capabilities of the spectrometers on our best current missions. High-resolution measurements by the dispersive grating spectrometers on \textit{Chandra} and \textit{XMM-Newton} (R$\sim$200--1000) are limited to point---or only slightly extended---sources, thus excluding targets such as galaxy clusters and entire supernova remnants. The CCD detectors on \textit{Chandra} and XMM only deliver moderate resolving power (R$\sim$50). As a result, many grand questions remain unsolved, such as: how did large scale baryonic structures form, and how do they evolve? How do black holes accrete and generate energetic winds and outflows, and how do they impact their surrounding environments? 

With future missions like \textit{XRISM} \cite{2018SPIE10699E..22T} and \textit{Athena} \cite{2013arXiv1306.2307N, barret18}, which will be equipped with high-resolution, non-dispersive micro-\\calorimeter imaging spectrometers with R$\sim$850 (\textit{XRISM}) and R$\sim$2400 (\textit{Athena}), we are finally poised to answer these questions. But we risk limiting the full scientific potential of these missions: \textbf{we currently lack an adequate understanding of many atomic processes behind the spectral features we will soon measure. The field of laboratory astrophysics, which comprises both theoretical and experimental studies of the underlying physics behind observable astrophysical processes, is thus central to the success of these missions.} In this White Paper, we highlight several science drivers for these future missions and identify specific laboratory astrophysics improvements required to~address~them. 

\vspace{-2ex}

\subsection*{Science Driver: AGN accretion, outflows, and feedback} \label{sec:scidrivers}
\vspace{-1ex}

Correlations between central black hole masses and host galaxy properties indicate that black holes and galaxies co-evolve through accretion and feedback \cite{2000ApJ...539L...9F, 2012ARA&A..50..455F}. A possible scenario is that AGN-driven winds overheat or sweep away the interstellar medium (ISM) from the galaxy bulge, quenching star formation and the AGN itself due to the lack of fuel for accretion. One key physical parameter required to validate this scenario is the amount of energy and metal contained in the wind to be deposited in the ISM, with the observational probe being Doppler-shifted atomic features in the UV and X-ray spectra. Observations with the \textit{Athena} X-IFU (X-Ray Integral Field Unit) will map the velocity flow to uncertainties of $\sim$20 km~s\textsuperscript{-1} \cite{barret16}, and track the metallicity of the hot gas in the AGN outflow and the atmospheres of the host galaxies \cite{2013arXiv1306.2330C}. But to correctly interpret these measurements and use them to determine densities, mass-loss rates, and momentum and energy fluxes, we require improved diagnostic theories with robust error estimates and experimental benchmarks. 

Spectral diagnostics of density, for example, involve ratios of lines which arise from metastable levels with differing critical densities. Fig.~1 shows one such example. Currently, very few of these transitions (on the order of 50\%) have sufficient atomic data such as transition energies, electron-impact excitation (EIE) strengths, and lifetimes. Furthermore, the ISM surrounding AGN is known to comprise a wide range of temperature components, which means that many ions exist in all possible charge states. Atomic data for M-shell Fe ions, whose lines can also be a probe for the ionization state, remain mostly unverified experimentally \cite{2001ApJ...563..497B} save a few exceptions \cite{2018ApJ...853...32B, 2001ApJ...557L..75B}. A 0.5 eV uncertainty on the energies of Fe lines, which is a typical theoretical value for L- and M-shell transitions in species of low charge states, will lead to $\gtrsim$100 km s\textsuperscript{-1} uncertainty on the feedback velocity. (Fig.~2 demonstrates the magnitude of this issue.) The outflowing hot gas is further subject to various population processes, in particular those of resonant excitation, dielectronic/radiative recombination (DR, RR), and inner-shell ionization. Many of these rates have $\sim$20\% uncertainty in theoretical calculations \cite{Gu, 2017JPhB...50f5203F}, which propagates to even larger uncertainties in the deduced chemical abundances, temperatures, and energies. 


Understanding the ionization mechanism for the gas surrounding AGN is also crucial for our understanding of feedback and the environment of AGN. In the traditional view, gas is photoionized by the active nucleus, but detailed spectroscopic and imaging analyses, particularly those tracing L-shell Fe emission, often reveal the need for a collisional component, related to star-forming activity or interaction with radio ejecta \cite{Bianchi2010a, Guainazzi2009, Braito2017, Kraemer2014}. However, uncertainties in the relevant atomic data \cite{Kraemer2014, Kinkhabwala2002, Kallman2014}, e.g., Fe fluorescence rates following inner-shell ionization, limits accurate estimates of the strength of this component. 

\vspace{-2ex}
\subsection*{Science Driver: Large-scale baryonic structure}
\vspace{-1ex}


The formation of groups and clusters of galaxies is a dynamic process: material builds by accretion and mergers, and potential energy is channelled through the intracluster medium (ICM) in the form of bulk motion and turbulence, eventually dissipating at smaller scales. The AGN in the brightest cluster galaxy drives gas motions in the central few $\sim$100 kpc of the ICM \cite{2014Natur.515...85Z}. These processes contribute to the overall virialization of the hot ICM within the halo potential well, and broaden and shift the emission line shapes from the region \cite{inogamov03}. Integrated over the line of sight, this might also result in distortions in the line profiles. 

Characterizing these emission lines allows us to probe the thermodynamics of the hot gas, and thus the formation of large-scale structure. However, to be able to correctly interpret these line shapes, we must disentangle the various interwoven processes that can also distort them. For example, DR satellite lines may blend with the parent line, charge exchange (CX) at interfaces between ionized and neutral gas may directly contribute to line flux, and resonant scattering (RS), which can also be a probe for gas velocities and anisotropies, may remove flux from certain lines. Comparisons between models and laboratory data for these processes often show significant differences (see, e.g., Fig.~3), requiring further attention.

 \begin{floatingfigure}[r]{3.35in}
\includegraphics[totalheight=2.2in]{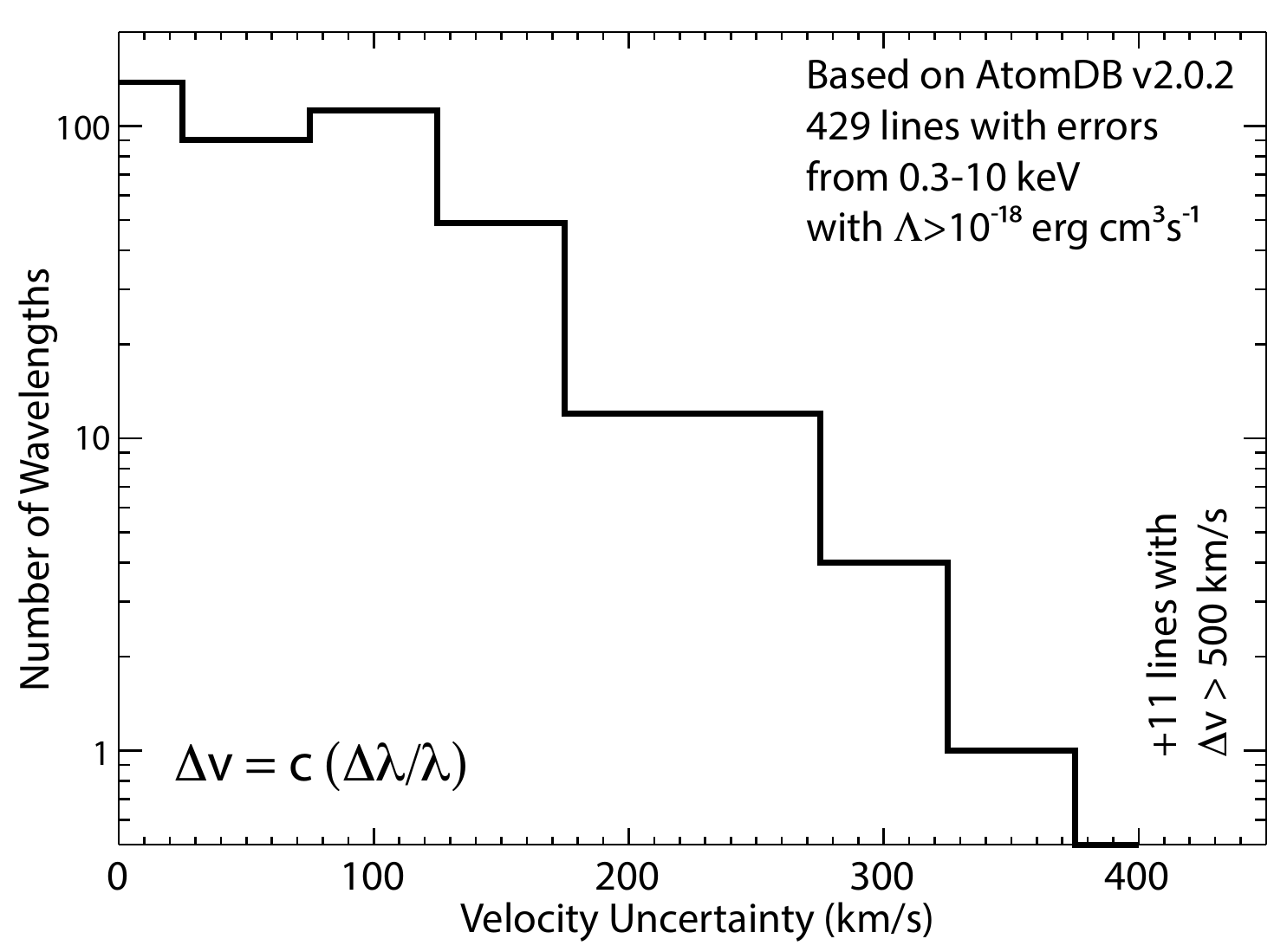}
\caption{\it Histogram of wavelength uncertainties for the strongest lines in the AtomDB v2.0.2 database. More than half have errors that will dominate the measured values from \textit{XRISM} and \textit{Athena}. From \cite{2014AAMOP..63..271S}.}
\end{floatingfigure}
 
Tracing the chemical evolution of the universe also holds clues to its formation history and the interconnected evolution of objects at various scales, from stars to galaxies and massive halos. This can be done by measuring the intensity of the brightest X-ray emission lines in the ICM of the most massive clusters to determine their abundances as a function of redshift. Observations with \textit{XMM-Newton} placed the first constraints on the relative contributions of core-collapse and type--Ia supernovae out to redshift $\sim$0.1 \cite{cheers}. Extending these studies to less abundant species and to larger redshift, as we expect to do with \textit{XRISM} and \textit{Athena}, is hampered by uncertainties in fundamental atomic quantities. For example, for many of the brightest emission lines in Ne-like and He-like Fe, radiative transfer effects have been shown to be significant \cite{2010SSRv..157..193C}. Unbiased measurements of the Fe abundance therefore have to rely on optically thin transitions, for which the EIE rates are less well known. This led to 16\% uncertainty on the derived Fe abundance of the Perseus cluster core through \textit{Hitomi} observations \cite{2018PASJ...70...12H}. Further, as we incorporate lower-mass systems with cooler plasmas, we will encounter limitations in the atomic data for energies and inner-shell excitation and ionization cross sections for lower charge state ions. 


Finally, an important aspect of large-scale structure formation involves ``missing" baryons that are predicted theoretically and expected to lie in the warm-hot intergalactic medium (WHIM), but that are difficult to detect. Any observation must subtract foreground contamination from galactic ISM, in particular inner-shell photoionization absorption lines from moderate to low charge ions. However, these do not have sufficiently benchmarked energies and cross sections. The source of discrepancies between the few existing laboratory measurements of transition energies and cross sections (e.g. \cite{2017AIPC.1811s0006G, 2017MNRAS.465.4690M, BESSY}) and \textit{Chandra} spectra remains a source of controversy.  

\vspace{-2ex}
\subsection*{Laboratory Astrophysics Requirements}
\vspace{-1ex}
Though here we only discuss two example science drivers, a multitude of other science cases are also affected by laboratory astrophysics needs, such as the study of solar and stellar winds, supernova remnants, dust in the ISM, and high density or magnetic field environments near black holes and neutron stars. To address the needs of the field, it is crucial to combine improved theoretical calculations with targeted laboratory measurements. Theory provides a framework for understanding, and modern computational techniques continue to improve the accuracy and consistency of many atomic calculations. Experiments are vital for interpreting real-life effects, identifying diagnostics, and benchmarking theory. This section highlights several specific requirements for atomic data relevant for future X-ray spectroscopy missions. 

%
{\bf Transition energies} in the rest-frame of the emitting ion are the most fundamental parameters required for any spectroscopic analysis. Their accuracy directly impacts our ability to infer plasma properties, such as density, temperature, and velocity, as shown above. To ensure that future line Doppler-shift measurements by, e.g., \textit{Athena} are not dominated by database uncertainties, we must have knowledge of line energies to better than 0.007\%. While H- and He-like ions have highly accurate line energies (to several parts per million for both theoretical calculations and laboratory measurements, with good agreement between the two \cite{2015PhRvA..91c2514B}), the situation becomes worse for lower charge state ions. For example, many inner-shell transitions from mid to low charge states are only known theoretically to $>$0.1\%, with a limited number of experimental benchmarks. Only theoretical energies are available for the 3$\rightarrow$2 inner-shell transitions of Fe (the ``M-shell unresolved transition array") often found in AGN spectra, themselves accurate at the 10-20\% level. The strongest DR satellite lines, which blend with the important ``triplet" lines of He-like ions used for electron density and temperature determinations (Fig.~1), are known to within $\sim$0.01\% (NIST), but few have been measured experimentally (e.g., \cite{2012CaJPh..90..351G}). Of the 429 strongest lines in the AtomDB X-ray line database with assigned errors, only $\sim$25\% of the lines are known to 20 km s\textsuperscript{-1} or better (see Fig. 2). Possibly more troublesome than having a large error, in many cases, there is no associated uncertainty in databases (see Fig.~5 in \cite{2014AAMOP..63..271S}). 


Laboratory energy measurements are resource- and time-intensive, but can yield promising results. Line energies have been measured with electron beam ion traps (EBITs, \cite{levine1988a}) with accuracies of about 0.01\% (e.g., \cite{brown1998, brown2002, gu2007, 2005ApJ...627.1066G, 2016ApJ...830...26H}). Experiments using trapped ions at synchrotron facilities have succeeded in measuring energies of inner-shell transitions of mid to low charge state ions at extremely high accuracies of $\sim$0.001\% (e.g., \cite{2018ApJ...853...32B, 2013PhRvL.111j3002R, BESSY}). Multiple theoretical methods for atomic calculations have also been used to obtain transition energies to 0.1--0.01\% accuracy.


{\bf Collisional (electron-impact) excitation (EIE)} contributes significantly to spectral lines in many astrophysical plasmas, and the resulting line ratios can be used for measurements of density, temperature, velocity, and elemental abundances, provided they are modeled correctly. Comparisons between atomic databases used in different codes revealed differences in the EIE collision strength for the Lyman-$\alpha$ transition in H-like Si, S, and Ar of up to 25\%, which leads to differences in the derived abundance of the same value \cite{2018PASJ...70...12H}. For the He-like Fe ``forbidden" line, this value increases to 42\%. Complete EIE models require large amounts of cross section data mostly provided through calculations (e.g., \cite{2014JPhCS.548a2010P, 2006ApJS..167..343B}), which can be benchmarked with experiments. EIE cross sections of specific transitions have been measured to $\sim$10\% accuracy with EBITs \cite{chen2008a} directly as a function of electron energy as well as for Maxwellian \cite{savin2000a} (or other) electron energy distributions \cite{chen2008a, 1995PhRvA..51.1214W, 2009JPhCS.163a2036T}.   


{\bf Photoexcitation/ionization and decay rates:} Accurate photoexcitation and ionization cross sections are necessary for understanding emission from optically thin, hot plasmas, such as those found in supernova remnants, stellar coronae, and the galactic ISM. Theoretical data exist (i.e.~\cite{1993A&AS...97..443K}), but their accuracy varies greatly and they require experimental validation, with the greatest needs being inner-shell transitions. Laboratory measurements of photoexcitation/ionization cross sections and oscillator strengths can be performed with an EBIT and a synchrotron or a free electron laser facility \cite{2010PhRvL.105r3001S, 2012Natur.492..225B}. In these experiments, one can simultaneously measure various decay channels after photoexcitation, e.g., Auger ionization and radiative rate branching ratios~\cite{PhysRevA.91.032502}. 

%

{\bf Dielectronic recombination (DR):} Accurate models of DR lines are critical for the proper interpretation of, e.g., the electron temperature in astrophysical plasmas \cite{1980RPPh...43..199D}. However, variations in DR rates of the strongest spectral lines between different databases may have led to large uncertainties in the derived abundance of the Perseus cluster from \textit{Hitomi} measurements \cite{2018PASJ...70...12H}. DR rates at low electron temperature are especially problematic, and require a combined theoretical/experimental approach. Measurements with ion storage rings can be used to deduce DR resonance positions and decay rates from spectra \cite{2005A&A...442..757F, 2004PhST..110..212G}, and EBIT measurements of resonances can be made as a function of quasi-Maxwellian electron temperature, allowing a comparison with spectral models \cite{gu2012, 2019ApJ...870...21B}.  

\begin{figure}
    \centering
    \includegraphics[totalheight=1.4in]{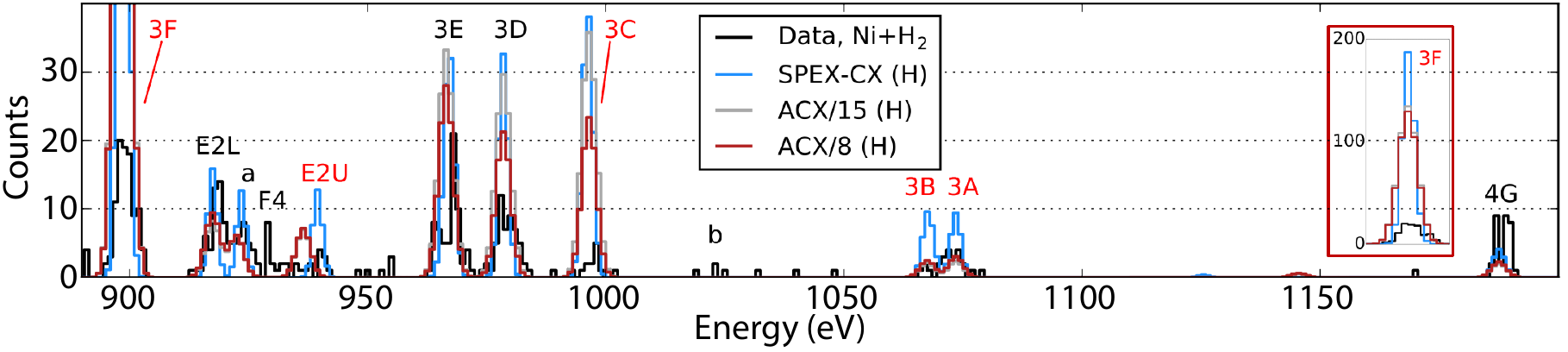}
    \setlength{\belowcaptionskip}{-10pt}
    \caption{\it Laboratory measurements of CX with $H_2$ on Ni L-shell ions compared with models, from \cite{2018ApJ...868L..17B}. Wide discrepancies exist between models and data, especially for the weaker lines.}
    \label{fig:my_label}
\end{figure}

{\bf Charge exchange recombination (CX)} is known to be the dominant X-ray emission mechanism within the Solar System (for further details, see the White Paper by Snios et al.), is a variable foreground to all observations \cite{2014Natur.512..171G}, and may also occur astrophysically \cite{2011ApJ...730...24K, 2014ApJ...787L..31C, 2007PASJ...59S.269T, 2015MNRAS.453.2480W}. However, as Fig. 3 demonstrates, models are often inaccurate or incomplete \cite{2018ApJ...868L..17B, 2010PhRvL.105f3201L, 2000PhRvL..85.5090B}, meaning that we risk inadequately removing foreground CX and assuming incorrect physical properties of our targets. Microcalorimeters coupled with EBITs (e.g. \cite{2014PhRvA..90e2723B}) or merged beam lines (e.g., \cite{2014PhRvA..89d2705F}) is a common method to benchmark model spectra at high resolution. Simultaneous cold-target recoil ion momentum spectroscopy (e.g. \cite{2010ApJ...716L..95A}) is also valuable for measuring $n$-resolved cross sections. A combination of these tools is also necessary to understand the effect of multi-electron processes, which can significantly alter the spectrum. 

%
%
\vspace{-2ex}
\subsection*{Looking Forward}
\vspace{-1ex}
The specific science cases and laboratory needs mentioned here give just a taste of the depth and importance of the field of laboratory astrophysics. We urge the Decadal Survey committee to follow all prioritized proposals from their top-level goals to the underlying atomic physics that is needed to understand them. Most tools and facilities needed to perform the recommended measurements and calculations exist, but increased funding is vital to ensure an adequate workforce at multiple career stages, the development of improved or new capabilities, and facility maintenance. We are about to enter into an era of high-resolution X-ray spectroscopy; we must ensure that we are able to fully reap its benefits.


\bibliography{bibliography}
\bibliographystyle{revtex}

\end{document}